\documentclass[aps,prl,twocolumn,showpacs,superscriptaddress]{revtex4}  

\usepackage{graphicx}
\usepackage{SIunits}
\usepackage{amsmath}
\begin{document}

\title{Rotational Quantum Friction}

\author{Rongkuo Zhao}
\email{r.zhao@imperial.ac.uk}
\affiliation{The Blackett Laboratory, Department of Physics, Imperial College London, London SW7 2AZ, United Kingdom}

\author{Alejandro Manjavacas}
\affiliation{IQFR - CSIC, Serrano 119, 28006 Madrid, Spain}

\author{F. Javier Garc\'{\i}a de Abajo}
\affiliation{IQFR - CSIC, Serrano 119, 28006 Madrid, Spain}
\author{J. B. Pendry}
\affiliation{The Blackett Laboratory, Department of Physics, Imperial College London, London SW7 2AZ, United Kingdom}

\date{\today}

\begin{abstract}
We investigate the frictional forces due to quantum fluctuations acting on a small sphere rotating near a surface. At zero temperature, we find the frictional force near a surface to be several orders of magnitude larger than that for the sphere rotating in vacuum.  For metallic materials with typical conductivity, quantum friction is maximized by  matching the frequency of rotation with the conductivity. Materials with poor conductivity are favored to obtain large quantum frictions. For semiconductor materials that are able to support surface plasmon polaritons, quantum friction can be further enhanced by several orders of magnitude due to the excitation of surface plasmon polaritons.
\end{abstract}


\maketitle
Fluctuation-induced electromagnetic forces, generally called Casimir forces \cite{Casimir-1948} which in the nonretarded limit are the van der Waals forces, dominate the interaction between nanostructures and cause permanent stiction in small devices such as micro- and nanoelectromechanical systems \cite{Serry,Buks}. Triggered by this urgent practical issue in nanoelectromechanical and the fast progress of force detection techniques,  experimental \cite{Mostepanenko-2009,Munday-2009,Chan-2010,Lamoreaux-2011a,Lamoreaux-2011b, Johnson-2011} and theoretical \cite{Zhao-2009,Johnson-2011} investigations on such \emph{static} fluctuation-induced electromagnetic forces between neutral bodies have experienced an extraordinary ``renaissance'' in the past few years.

Consider two surfaces separated by a finite distance.  Quantum fluctuations create instantaneous charges on the surfaces. If the surfaces are in relative parallel motion, induced image charges lag behind and tend to pull the fluctuating charges back. This lateral dynamical fluctuation-induced electromagnetic interaction yields a noncontact friction between two perfectly smooth featureless dielectric plates. The electrical resistance of the material dissipates the frictional work. This lateral friction is called quantum friction, which was first studied in detail by Pendry in 1997 \cite{Pendry-1997,Pendry-1998}.  Volokitin and Persson then further studied the quantum friction between two parallel surfaces and that of a small sphere (or a neutral atom) moving parallel to a surface \cite{Persson-2007}. Despite the mounting theoretical evidence of the existence of these types of forces, recent work [14] has questioned the existence of quantum friction at absolute zero temperature between two parallel surfaces, leading to heated debates \cite{Pendry-2009,Persson-2011c}.  Note the distinction between quantum friction between two perfectly smooth surfaces and the lateral Casimir force between a noncontacting corrugated plate and a corrugated cylinder \cite{Golestanian-2007}.

In this Letter, we investigate the quantum friction acting on a rotating small neutral sphere positioned close to a surface at zero and nonzero temperatures. As in a previously reported study [18] for an isolated sphere rotating in vacuum, our calculations unambiguously confirm the existence of quantum friction at absolute zero temperature. Because of the huge local density of electromagnetic states (LDOS) near a surface \cite{Greffet-2003}, quantum friction is enhanced by several orders of magnitude compared with that for a sphere rotating in free space studied in Ref. \cite{Abajo-2010}. The relation between quantum friction and the conductivity is thoroughly investigated. In particular, we present calculations for a realistic semiconductor material, indium antimonide (InSb), which can support the excitation of surface plasmon polaritons (SPPs) and lead to larger LDOS near a surface at  the SPP frequency. The larger LDOS can further enhance quantum friction by several orders of magnitude. This enhancement opens up the possibility of experimental verification.

\begin{figure}[b]
\centering{\includegraphics[angle=0, width=8cm]{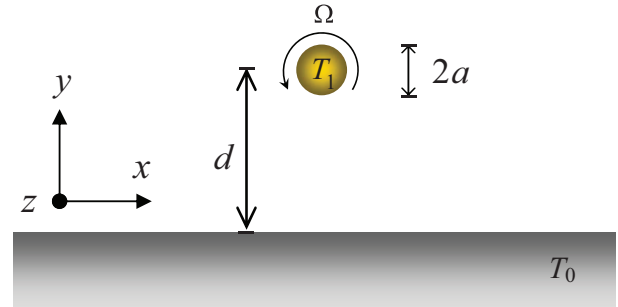}}
\caption{(color online) Sketch of a spherical particle rotating near a surface.  A spherical particle at temperature $T_1$ with radius $a$ rotates along the $z$ axis with frequency $\Omega$ and is positioned at a distance $d$ from  a semi-infinite homogeneous isotropic medium at temperature $T_0$.}
 \label{Fig:schetch}
\end{figure}

We focus on a spherical homogeneous particle of radius $a$ placed in vacuum at a distance $d$ from a semi-infinite homogeneous medium, rotating at frequency $\Omega$ around a direction parallel to the surface of the latter, as shown in Fig. 1. The particle is at temperature $T_1$, while the planar surface and the vacuum are both at temperature $T_0$. We work within the electrostatic limit, we neglect multiple scattering between the particle and the surface, and we describe the particle through its electrical polarizability $\alpha(\omega)=a^3[\epsilon(\omega)-1]/[\epsilon(\omega)+2]$, which is written in terms of its dielectric function $\epsilon(\omega)$. These assumptions are valid for $a\ll d$ and $d$ much smaller than both $c/\Omega$ and $\hbar c/k_BT_j$.

By symmetry, the friction torque acting on the sphere is along the $z$ axis and given by a spectral integral (see \cite{Abajo-2010} for a comprehensive derivation or \cite{SupplMat} for a simple derivation)
\begin{equation}\label{torque}
M=-\frac{2\hbar}{\pi}\int^\infty_{-\infty}\Gamma(\omega)d\omega,
\end{equation}
where $\Gamma(\omega)=[n_1(\omega-\Omega)-n_0(\omega)]\text{Im}\{\alpha(\omega-\Omega)\}\text{Im}\{\bar{G}(\omega)\}$ is the spectral distribution of the torque \cite{multiscattering}. $n_j(\omega)=[\text{exp}(\hbar\omega/k_BT_j)-1]^{-1}$ is the Bose-Einstein distribution function at temperature $T_j$.  Here, $\bar{G}(\omega)= [G_{xx}(\omega)+G_{yy}(\omega)]/2$, where $G_{ij}$ is the electromagnetic Green tensor connecting the fluctuating dipole moment $\textbf{p}^\text{fl}(\omega)$ to the induced electromagnetic field $\textbf{E}^\text{ind}(\omega)$  at the position of the sphere $\textbf{r}_0$, i.e., $\textbf{E}^\text{ind}(\omega)=\textbf{G}(\textbf{r}_0,\textbf{r}_0,\omega) \cdot \textbf{p}^\text{fl}(\omega)$ \cite{Abajo-2010}. Neglecting a minor contribution from the vacuum, the Green tensor components near the planar surface reduce in the electrostatic limit to $G_{xx}(\omega)=G_{zz}(\omega)=G_{yy}(\omega)/2=(2d)^{-3}[\epsilon(\omega)-1]/[\epsilon(\omega)+1]$. The heat transfer power $P_{1\rightarrow0}$ from the sphere to the surface is similarly given by \cite{Abajo-2010, SupplMat, multiscattering}
\begin{equation}\label{heattransfer}
P_{1\rightarrow 0}=\frac{2\hbar}{\pi}\int^\infty_{-\infty}\omega\Gamma(\omega)d\omega+\frac{\hbar}{\pi}\int^\infty_{-\infty}\omega\Gamma_0(\omega)d\omega,
\end{equation}
where $\Gamma_0(\omega)=[n_1(\omega)-n_0(\omega)]\text{Im}\{\alpha(\omega)\}\text{Im}\{G_{zz}(\omega)\}$.
The second integral in Eq.~(\ref{heattransfer}) vanishes when the sphere and the surface are at the same temperature. Incidentally, when the sphere rotates around an axis perpendicular to the planar surface (i.e., the $y$ axis in Fig. 1), $G_{yy}$  and $G_{zz}$  are swapped in Eqs. (1) and (2). The dominant frequency range contributing to Eqs. (\ref{torque}) and (\ref{heattransfer}) is generally determined by the prefactor $[n_1(\omega-\Omega)-n_0(\omega)]$. The range is below $\Omega$ and $k_B T_j/\hbar$ limited by both kinematical and thermal frequencies \cite{Jentschura-2012}. At zero temperature, $[n_1(\omega-\Omega)-n_0(\omega)]$ becomes a step function taking the value $-1$ in the frequency window [0,~$\Omega$]. In this Letter, we neglect contributions from fluctuating magnetic dipole moments and fluctuating magnetic fields. These contributions become dominant when $d\sigma_0/c\gg1$ \cite{Greffet-2003} or $a\sigma_0/c\gg1$ \cite{Abajo-2010}.  However, materials with low conductivity lead to large quantum friction (see below)  and then magnetic contributions can be safely neglected.

According to the fluctuation dissipation theorem, $\text{Im}\{\alpha(\omega)\}$ is proportional to the magnitude of the dipole moment fluctuation \cite{Abajo-2010}, while $\text{Im}\{G_{ij}(\omega)\}$ is proportional to the magnitude of the electric field fluctuation, which is in turn scaling with the LDOS near the surface \cite{Greffet-2003}. These quantities are solely determined by the permittivity of the material.  We now  study the quantum friction using different materials. For simplicity, we assume the sphere and the surface to be made of the same material.

\emph{Metallic materials}.---We consider a material with high conductivity $\sigma_0$, the optical response of which is well described by the Drude permittivity $\epsilon(\omega)=1+i4\pi\sigma_0/\omega$ at low frequencies. If the relevant frequencies $k_BT_j/\hbar$ and $\Omega$ are much smaller than the conductivity (we use Gaussian units, in which the conductivity has dimension of frequency), we have
\begin{align}\label{alpha_highconductivity}
\text{Im}\{\frac{\epsilon-1}{\epsilon+2}\}&\simeq 3\omega /4\pi\sigma_0,\\
\text{Im}\{\frac{\epsilon-1}{\epsilon+1}\}&\simeq \omega /2\pi\sigma_0,
\end{align}
and consequently we have closed-form expressions for Eqs. (\ref{torque}) and (\ref{heattransfer}):
\begin{equation}\label{M_highconductivity}
M=-\frac{3\hbar}{256\pi^3\sigma_0^2}\frac{a^3}{d^3}(\theta_1^2+\theta_0^2+2\Omega^2)\Omega,
\end{equation}
\begin{equation}\label{P_highconductivity}
P_{1\rightarrow0}=\frac{\hbar}{128\pi^3\sigma_0^2}\frac{a^3}{d^3}[\frac{\theta_1^4-\theta_0^4}{5}+\frac{3}{2}(\theta_1^2\Omega^2+\Omega^4)],
\end{equation}
where $\theta_j=2\pi k_BT_j/\hbar$. Equation (\ref{M_highconductivity})  verifies the existence of quantum friction as $M\propto\Omega^3$ at absolute zero temperature, to be compared with $M\propto\Omega^5$ for the sphere rotating in free space \cite{Abajo-2010}.  The torque due to quantum friction is an odd function of the rotation frequency $\Omega$ as a consequence of causality (notice that the $\text{Im}\{\dots\}$ factors in $\Gamma$ are odd functions of $\omega$), so it always results in mechanical stopping regardless of the sign of $\Omega$.  The mechanical energy dissipation $-M\Omega$ is not equal to the heat transfer power $P_{1\rightarrow0}$ from the sphere to the surface. The remaining part of the energy, $P_{\text{abs}}=-M\Omega-P_{1\rightarrow0}$, is absorbed by the particle in the form of thermal heating. $P_{\text{abs}}=0$ determines the equilibrium temperature condition, which is $\frac{\theta_1}{\theta_0}=[1+\frac{15}{2}(\frac{\Omega}{\theta_0})^2+\frac{15}{2}(\frac{\Omega}{\theta_0})^4]^{1/4}$. Quantum friction near the surface produces particle heating similar to conventional friction, which is different from the result for the sphere rotating in free space, for which cooling of the particle has been claimed to be possible [18].  The heat transfer power is positively tuned by the rotation frequency. It is nonzero (as $P_{1\rightarrow0}\propto\Omega^4$) even at absolute zero temperature, which shows spontaneous emission due to rotation \cite{Abajo-2010,Kardar-2012}. Equation (\ref{P_highconductivity}) agrees with the previous result in Ref. \cite{Pendry-1999} when $\Omega=0$.

In the limit of low rotation frequencies ($\Omega\ll\theta_j\ll\sigma_0$), we can assume negligible heating (i.e., $T_0=T_1=T$) so that the torque becomes proportional to $\Omega$, $M\simeq-\beta\Omega$, where $\beta=3k_B^2T^2a^3/32\hbar\pi\sigma_0^2d^3$. This leads to a time-dependent rotation velocity $\Omega(t)=\Omega(0)\text{exp}(-t/\tau)$ characterized by a stopping time $\tau=I/\beta$, where $I=8\rho\pi a^5/15$ is the moment of inertia of the sphere and $\rho$ is its mass density. More precisely, $\tau=256\rho\hbar\pi^2 a^2 d^3 \sigma_0^2/45k_B^2T^2.$

\begin{figure}[t]
\centering{\includegraphics[angle=0, width=8.1cm]{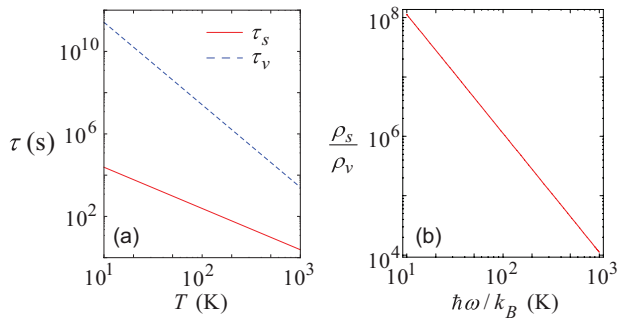}}
\caption{(color online) (a) Characteristic stopping time of a graphite sphere rotating close to a graphite surface (solid line, $\tau_s$) and in free space \cite{Abajo-2010} (dashed line, $\tau_v$)  as a function of temperature.  (b) Ratio of the LDOS near a surface, $\rho_s$, to that in free space, $\rho_v$, at low frequencies. The frequency is converted to temperature as $\hbar \omega/k_B$. We take $a=5$ nm and $d=30$ nm. The conductivity of graphite is $\sigma_0=2.1 \times 10^{14}~ \text{s}^{-1}$ as used in Ref. \cite{Abajo-2010}.}
\label{Fig:stoppingtime}
\end{figure}

By comparing the above result with the characteristic stopping time of a sphere rotating in vacuum (dashed line in Fig.~\ref{Fig:stoppingtime}(a)) studied in Ref. \cite{Abajo-2010}, the stopping time close to a surface (solid line in Fig.~\ref{Fig:stoppingtime}(a)) is several orders of magnitude smaller. For instance, at room temperature, the stopping time decreases from 4 days to 30 s, which is feasible to measure in an experiment. This enhancement of quantum friction is attributed to the LDOS close to the surface, which is several orders of magnitude larger than that in free space. In the low frequency limit, the ratio of the LDOS near a surface, $\rho_s$, to that in vacuum, $\rho_v$, is $9 c^3/64\pi d^3\omega^2\sigma$ \cite{Greffet-2003}, much larger than 1, as shown in Fig.~\ref{Fig:stoppingtime}(b). This ratio explains the enhancement of quantum friction.

Equation (\ref{M_highconductivity}) shows that lower conductivity leads to shorter stopping times. However, it is valid only when the relevant frequencies are much smaller than the conductivity. For high rotation speed or high temperature compared with the conductivity, the dominant frequency range contributing to Eq.~(1) extends into the frequency region above $\sigma_0$. Therefore, $\omega/\sigma_0$ is no longer small and the approximated expressions Eqs.~(3) and (4) are not valid. Instead, we retain the full $\omega$ dependence of $\epsilon$ in the numerical evaluation of Eq.~(\ref{torque}), which we rewrite as \begin{equation*}  M=-3\hbar a^3 J/8\pi d^3. \end{equation*}   $J$ is evaluated numerically.

\begin{figure}[t]
\centering{\includegraphics[angle=0, width=7cm]{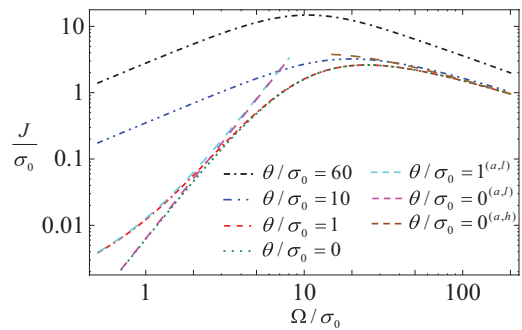}}
 \caption{(color online) $J$ versus the rotation frequency at different temperatures, normalized to $\sigma_0$. The dashed curves are the asymptotic results in the low and high rotation frequency limits. The superscripts $(a,l)$ and $(a,h)$ indicate the asymptotic curves in the low and high rotation frequency limits.}
 \label{Fig:J_Conductor2D}
\end{figure}
We consider two interesting limiting cases (but with the temperature not too high, see dashed curve in Fig.~\ref{Fig:J_Conductor2D}): In the $\Omega\ll\sigma_0$ limit, the asymptotic expression is Eq.~(\ref{M_highconductivity}); in the $\Omega\gg\sigma_0$ limit, we find
\begin{equation*}
J_{\Omega\gg\sigma_0}=\frac{8\pi^2\sigma}{3}[\frac{2\ln(\Omega/\sigma_0)-\ln(8\pi^2/3)}{\Omega/\sigma}],
\end{equation*}
which is a decreasing function of the rotation frequency. Therefore, quantum friction reaches its maximum at an intermediate rotation frequency.

The other curves in Fig.~\ref{Fig:J_Conductor2D} show the full numerical results at different temperatures. At low temperatures, frictions at low rotation frequencies are significantly influenced by the temperature. They converge when the rotation frequency is larger than a certain value. This value is proportional to the temperature.  Such behavior persists until the temperature $\theta$ is sufficiently larger than the conductivity. As a consequence, the frequencies of maximum friction are different at different temperatures. At zero temperature, $J$ reaches a maximum of $2.6088 \sigma_0$ when $\Omega=24.7679\sigma_0$. At high temperatures, the maximum of $J$ is $0.25\theta$ when $\Omega=10\pi\sigma_0/3$, so the maximum quantum friction is

\begin{equation*}
M_\text{max}=-\frac{3a^3}{16d^3}k_B T,
\end{equation*}
which depends on temperature only linearly. Therefore, given a fixed conductivity, quantum friction first increases and then decreases with increasing spin speed. The maximum is obtained at a speed between $10\pi\sigma_0/3$ and $24.7679\sigma_0$ depending on the temperature. A three-dimensional plot of $J$ versus the rotation frequency and temperature is shown in Fig.~S1 \cite{SupplMat}.

Given a fixed rotation frequency, we observe similar behavior (Fig.~S2 \cite{SupplMat}): $J$ first increases and then decreases with increasing conductivity. At zero temperature, $J$ reaches a maximum of $0.1615\Omega$ when $\sigma_0=0.0883\Omega$; at high temperatures, $J$ reaches a maximum of $0.25\theta$ when $\sigma_0=3\Omega/10\pi$; at intermediate temperatures, $J$ reaches a maximum when $\sigma_0$ is between  $0.0883\Omega$ and $3\Omega/10\pi$.

To maximize the quantum friction, one needs to match the rotation frequency with the conductivity. Note, however, that for the graphite with $\sigma_0=2.1 \times 10^{14}~ \text{s}^{-1}$, it is extremely difficult for a macroscopic sphere (10 nm) to spin so fast. Even for diatomic molecules, the rotational constant typically varies from about $1.5\times 10^{9}$ to $6\times 10^{11} ~ \text{s}^{-1}$. Therefore, considering these experimental challenges, it can be more advantageous to utilize materials with poorer conductivity such as semiconductors. For instance, the conductivities of germanium and silicon are in the $10^{8}$ to $10^{14} ~ \text{s}^{-1}$ range, depending on the impurity concentration \cite{Irvin-1968}.

\emph{Realistic semiconductor material}.---In the following, we study quantum friction using a realistic semiconductor material, InSb. Its optical permittivity can be described by a Lorentz model adding a Drude term \cite{Palik}:
\begin{equation}\label{epsilon_InSb}
\epsilon=\epsilon_\infty[1+\frac{\omega_L^2-\omega_T^2}{\omega_T^2-\omega^2-i\Gamma\omega}-\frac{\omega_p^2}{\omega(\omega+i\gamma)}],
\end{equation}
where  $\epsilon_\infty=15.68$, $\omega_T=179.1~\text{cm}^{-1}$, $\omega_L=190.4~\text{cm}^{-1}$, $\Gamma=2.86~\text{cm}^{-1}$, $\omega_p=81.0~\text{cm}^{-1}$, and $\gamma=10.7~\text{cm}^{-1}$. The surface of InSb can support SPPs when Re$\{\epsilon\}=-1$ at $\omega_1=2.18$ THz and $\omega_2=$ 5.75 THz, which results in huge peaks in the LDOS \cite{Greffet-2003}. Moreover, the spherical particle can support localized SPPs when Re$\{\epsilon\}=-2$ at $\omega_1^L=2.12$ THz and $\omega_2^L=$ 5.73 THz, which induces a strong electromagnetic response on the spherical particle. $\omega_1^L$ and $\omega_2^L$ are very close to $\omega_1$ and $\omega_2$, respectively, and we assume that they are the same in the following discussion. Therefore, we could obtain much larger quantum friction by taking advantage of the huge LDOS of SPPs at the surface and the strong electromagnetic response of the localized SPPs on the sphere.

\begin{figure}[b]
 \centering{\includegraphics[angle=0, width=8.4cm]{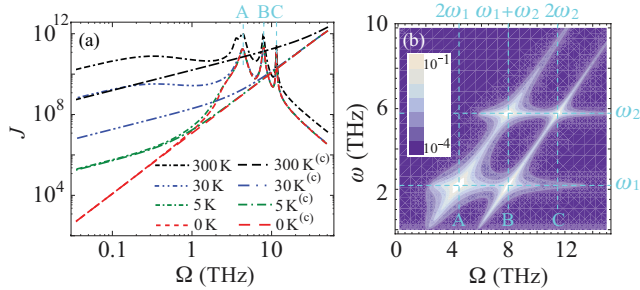}}
 \caption{(color online) (a) $J$ versus the rotation frequency at different temperatures for InSb described by Eq.~(\ref{epsilon_InSb}) (short dashes) and a metallic material with an equivalent conductivity $\sigma_0=\epsilon_\infty \omega_p^2/4 \pi\gamma$ (long dashes). The peaks are marked by A, B, and C. (b) Integrand of $J$, $[n(\omega-\Omega)-n(\omega)] \text{Im}\{[\epsilon(\omega-\Omega)-1]/[\epsilon(\omega-\Omega)+2]\} \text{Im}\{[\epsilon(\omega)-1]/[\epsilon(\omega)+1]\}$, versus the rotation frequency and optical frequency at T=0. A, B, and C correspond to the frequencies $2\omega_1$, $\omega_1+\omega_2$, and $2\omega_2$, respectively. The plot in (b) is saturated below $10^{-4}$ in order to avoid divergent values in the log scale.}
 \label{Fig:J_InSb}
\end{figure}

In Fig.~\ref{Fig:J_InSb}(a), we compare quantum frictions at different temperatures for InSb (short dashes) and a metallic material (long dashes) described by the simplified Drude model  with the permittivity function  $\epsilon=1+i4 \pi\sigma_0/\omega$ and an equivalent conductivity $\sigma_0=\epsilon_\infty \omega_p^2/4 \pi\gamma$.  InSb can support both propagating and localized SPP excitations, while the equivalent Drude metallic material with the equivalent conductivity cannot. At low temperatures ($\text{T} < 5\text{~K}$) and low rotation frequencies ($\Omega<1~\text{THz}$), the results for InSb are in excellent agreement with those for the equivalent metallic material, which confirms the validity of the simplified Drude model in these limits.   However, at high temperatures ($\text{T} >5\text{~K}$) or high rotation frequencies ($\Omega>1~\text{THz}$) (i.e., the relevant  frequencies  $k_BT/\hbar$ or $\Omega$  are close to or larger than the SPP frequencies $\omega_1$ and $\omega_2$), the results for InSb are several orders of magnitude larger than those for the equivalent metallic material. Interestingly, if the temperature frequency $k_BT/\hbar$ is high enough to be comparable with the SPP frequencies, this enhancement persists even when the rotation frequency is much lower than the SPP frequencies. This temperature is not high and can be easily achieved in experiments (see the dash-dot-dotted and dash-dash-dotted curves in Fig.~\ref{Fig:J_InSb}(a)). For a rotation frequency much larger than the SPP frequencies, the results for InSb decrease much faster than those for the equivalent metallic material.

The three peaks marked by A, B, and C in Fig.~\ref{Fig:J_InSb}(a) are due to the SPP and localized SPP excitations. As shown in Fig.~\ref{Fig:J_InSb}(b), at T=0,  the integrand of $J$, $[n(\omega-\Omega)-n(\omega)] \text{Im}\{[\epsilon(\omega-\Omega)-1]/[\epsilon(\omega-\Omega)+2]\} \text{Im}\{[\epsilon(\omega)-1]/[\epsilon(\omega)+1]\}$, versus the rotation frequency and optical frequency  has four ``singular'' lines: $\omega=\omega_1$, $\omega=\omega_2$, $\omega=\Omega-\omega_1$, and $\omega=\Omega-\omega_2$ determined by the SPP and localized SPP frequencies. Four singular spots at the intersections of the four singular lines correspond to those three peaks at $2\omega_1$, $\omega_1+\omega_2$, and $2\omega_2$. This indicates that the SPP excitations can further enhance quantum friction by several orders of magnitude even at T=0 once the rotation frequency $\Omega$ reaches the plasmon resonance frequencies. It also indicates that insulators such as SiC or SiO$_2$, which exhibit phonon polaritons in the infrared, cannot produce considerable quantum friction at T=0 until the spinning speed reaches the polariton frequency.

In conclusion, we have made a theoretical investigation of the quantum friction acting on a sphere rotating near a surface. The existence of quantum friction at absolute zero temperature is confirmed. Because of the huge density of electromagnetic states close to the surface, the friction near the surface is several orders of magnitude larger than that in free space. For metallic materials, maximizing the quantum friction requires matching the rotation frequency with the conductivity. Materials with poor conductivity are thus favored for plausible experiments. Moreover, quantum friction can be further enhanced by several orders of magnitude when the characteristic temperature frequency $k_BT/\hbar$ or the rotation frequency is high enough to reach the surface plasmon resonance frequencies in some semiconductor materials. This significant enhancement of quantum friction opens up the possibility of experimental realizations of these phenomena. The challenges for future experiments are to enable particles to rotate at a high speed and close to a surface.

This work is supported by the European Community
project PHOME (Contract No. 213390), the Leverhulme Trust, and the Spanish MEC (Consolider NanoLight.es). A.M. acknowledges financial support through FPU from ME.

\begin{widetext}

\newpage

\section{Supplemental Material: Derivations of the torque and the heat transfer power}

\setcounter{equation}{0}

In this Supplemental Material we follow a similar method and the same notations as in Ref. \cite{Abajo-2010}. Our problem consists of a semi-infinite medium, a sphere above it, and a semi-infinite free space. In the electrostatic limit, we neglect the contribution from the free space and consider the interaction between the sphere and the semi-infinite medium only. The instantaneous fluctuating charges on the surface of the semi-infinite medium generate the electromagnetic field $\textbf{E}^\text{fl}(\omega)$ near the surface which induces a dipole moment $\textbf{p}^\text{ind}(\omega)=\alpha(\omega) \textbf{E}^\text{fl}(\omega)$ on the sphere; The fluctuating charges on the sphere generate the electric dipole moment $\textbf{p}^\text{fl}(\omega)$ which induces the electromagnetic field $\textbf{E}^\text{ind}(\omega)=\textbf{G}(\omega) \textbf{p}^\text{fl}(\omega)$ near the surface.

The friction torque given by $\textbf{p}\times\textbf{E}$ can be separated into two  parts: (1) $M_p$ originates from the dipole moment fluctuation; (2) $M_E$ originates from the electrical field fluctuation.
\begin{align*}
M_p=<p_x^\text{fl}(\omega)E_y^\text{ind}(\omega)-p_y^\text{fl}(\omega)E_x^\text{ind}(\omega)>,\\
M_E=<p_x^\text{ind}(\omega)E_y^\text{fl}(\omega)-p_y^\text{ind}(\omega)E_x^\text{fl}(\omega)>.
\end{align*}

Now we calculate the first part. Assume that we have a fluctuating dipole moment $\textbf{p}^\text{fl}(\omega,t)$ on the $x-y$ plane rotating along the $z$ axis near the surface at a speed $\Omega$. In the rest frame, this dipole moment can be decomposed as
\begin{align*}
p_x^\text{fl}(\omega,t)=p_0^\text{fl}\cos(\omega t+\gamma)\cos(\Omega t+\beta),\\
p_y^\text{fl}(\omega,t)=p_0^\text{fl}\cos(\omega t+\gamma)\sin(\Omega t+\beta),
\end{align*}
where $p_0^\text{fl}$ is the magnitude of the fluctuating dipole moment; $\gamma$ and $\beta$ are the initial phases. Rewriting them in the exponential forms, we have
\begin{align*}
p_x^\text{fl}(\omega,t)=\frac{p_0^\text{fl}}{4}[e^{-i\omega^+t-i\gamma^+}+e^{+i\omega^+t+i\gamma^+}+e^{-i\omega^-t-i\gamma^-}+e^{+i\omega^-t+i\gamma^-}],\\
p_y^\text{fl}(\omega,t)=\frac{ip_0^\text{fl}}{4}[e^{-i\omega^+t-i\gamma^+}-e^{+i\omega^+t+i\gamma^+}-e^{-i\omega^-t-i\gamma^-}+e^{+i\omega^-t+i\gamma^-}],
\end{align*}
where $\omega^\pm=\omega\pm\Omega$ and $\gamma^\pm=\gamma\pm\beta$. Using the relation $\textbf{E}^\text{ind}(\omega)=\textbf{G}(\omega) \textbf{p}^\text{fl}(\omega)$, we have
\begin{align*}
E_x^\text{ind}(\omega,t)=\frac{p_0^\text{fl}}{4}[G_{xx}(\omega^+)e^{-i\omega^+t-i\gamma^+}+G_{xx}(-\omega^+)e^{+i\omega^+t+i\gamma^+}+G_{xx}(\omega^-)e^{-i\omega^-t-i\gamma^-}+G_{xx}(-\omega^-)e^{+i\omega^-t+i\gamma^-}],\\
E_y^\text{ind}(\omega,t)=\frac{ip_0^\text{fl}}{4}[G_{yy}(\omega^+)e^{-i\omega^+t-i\gamma^+}-G_{yy}(-\omega^+)e^{+i\omega^+t+i\gamma^+}-G_{yy}(\omega^-)e^{-i\omega^-t-i\gamma^-}+G_{yy}(-\omega^-)e^{+i\omega^-t+i\gamma^-}].
\end{align*}
Then omitting the oscillation term, we have
\begin{align*}
M_p=\frac{-i(p_0^\text{fl})^2}{16}\{[G_{xx}(-\omega^+)-G_{xx}(\omega^+)+G_{xx}(\omega^-)-G_{xx}(-\omega^-)]+[G_{yy}(-\omega^+)-G_{yy}(\omega^+)+G_{yy}(\omega^-)-G_{yy}(-\omega^-)]\}.
\end{align*}
Using the causality property of the Green tensor $\textbf{G}(-\omega)=\textbf{G}^*(\omega)$ and rewriting $\bar{G}(\omega)=[G_{xx}(\omega)+G_{yy}(\omega)]/2$, we have
\begin{equation}\label{sM_pp}
M_p=\frac{(p_0^\text{fl})^2}{4}[\text{Im}\{\bar{G}(\omega^-)\}-\text{Im}\{\bar{G}(\omega^+)\}].
\end{equation}
The final result Eq. (\ref{sM_pp}) does not depend on the initial phases. Taking into account the dipole moment fluctuating along another direction (perpendicular to what we have considered above), the torque should be multiplied by 2. Equation (\ref{sM_pp}) tells us that the torque originates from the dispersion of the imaginary part of Green tensor and the frequency splitting due to  rotation. The frequency splitting in the rotation system is similar to the Doppler shift of the frequency in parallel surface system studied in Pendry's original paper in 1997. The physical meaning of the imaginary part of Green tensor is associated with the local density of electromagnetic states.

Then we calculate the second part.  Assume that we have a fluctuating electric field  $E^\text{fl}_{0x}\cos(\omega t+\gamma)$ in the rest frame along the $x$ axis. In the rotating frame, this electric field can be decomposed as
\begin{align}
E_{x-\text{rot}}^\text{fl}(\omega,t)=E^\text{fl}_\text{0x}\cos(\omega t+\gamma)\cos(\Omega t+\beta),\\
E_{y-\text{rot}}^\text{fl}(\omega,t)=-E^\text{fl}_\text{0x}\cos(\omega t+\gamma)\sin(\Omega t+\beta),
\end{align}
where $\gamma$ and $\beta$ are the initial phases.  Note the minus sign in Eq. (3). If we are in the rotating frame, the rest frame rotates in the opposite direction. Rewriting them in the exponential forms, we have
\begin{align*}
E_{x-\text{rot}}^\text{fl}(\omega,t)=\frac{E^\text{fl}_\text{0x}}{4}[e^{-i\omega^+t-i\gamma^+}+e^{+i\omega^+t+i\gamma^+}+e^{-i\omega^-t-i\gamma^-}+e^{+i\omega^-t+i\gamma^-}],\\
E_{y-\text{rot}}^\text{fl}(\omega,t)=\frac{-iE^\text{fl}_\text{0x}}{4}[e^{-i\omega^+t-i\gamma^+}-e^{+i\omega^+t+i\gamma^+}-e^{-i\omega^-t-i\gamma^-}+e^{+i\omega^-t+i\gamma^-}].
\end{align*}
Using the relation $\textbf{p}^\text{ind}(\omega)=\alpha(\omega) \textbf{E}^\text{fl}(\omega)$, we have
\begin{align*}
p_{x-\text{rot}}^\text{ind}(\omega,t)=\frac{E^\text{fl}_{0x}}{4}[\alpha_{xx}(\omega^+)e^{-i\omega^+t-i\gamma^+}+\alpha_{xx}(-\omega^+)e^{+i\omega^+t+i\gamma^+}+\alpha_{xx}(\omega^-)e^{-i\omega^-t-i\gamma^-}+\alpha_{xx}(-\omega^-)e^{+i\omega^-t+i\gamma^-}],\\
p_{y-\text{rot}}^\text{ind}(\omega,t)=\frac{-iE^\text{fl}_{0x}}{4}[\alpha_{yy}(\omega^+)e^{-i\omega^+t-i\gamma^+}-\alpha_{yy}(-\omega^+)e^{+i\omega^+t+i\gamma^+}-\alpha_{yy}(\omega^-)e^{-i\omega^-t-i\gamma^-}+\alpha_{yy}(-\omega^-)e^{+i\omega^-t+i\gamma^-}].
\end{align*}
Then we can calculate the torque either in the rotating frame or in the rest frame because the torque is conserved in either system. However, the radiation power is not conserved. we have to go to the rest frame to calculate the radiation power.

Now we choose the rotating frame to calculate the torque since it is easier. Omitting the index of $\alpha$ for an isotropic sphere and the oscillation terms, we have
\begin{align*}
M_E=\frac{-i(E_{0x}^\text{fl})^2}{8}[\alpha(-\omega^+)-\alpha(\omega^+)+\alpha(\omega^-)-\alpha(-\omega^-)].
\end{align*}
Using the causality property of the polarizability $\alpha(-\omega)=\alpha^*(\omega)$, we have
\begin{equation*}
M_E=\frac{(E_{0x}^\text{fl})^2}{4}[\text{Im}\{\alpha(\omega^-)\}-\text{Im}\{\alpha(\omega^+)\}].
\end{equation*}
Taking into account the fluctuating electric field  along the $y$ axis, we then have
\begin{equation}\label{sM_EE}
M_E=\frac{(E_{0x}^\text{fl})^2+(E_{0y}^\text{fl})^2}{4}[\text{Im}\{\alpha(\omega^-)\}-\text{Im}\{\alpha(\omega^+)\}].
\end{equation}
Similarly, Eq. (\ref{sM_EE}) tells us that the torque originates from the dispersion of the imaginary part of the polarizability and the frequency splitting due to rotation.

Now the next question arises, what are $E_{0j}^\text{fl}$ and $p_{0}^\text{fl}$?  According to the fluctuation dissipation theorem \cite{Abajo-2010},  $(E_{0x}^\text{fl})^2$ and $(p_{0}^\text{fl})^2$ are given by the imaginary parts of the Green tensor and the dipole moment polarizability respectively:

\begin{equation}\label{sE}
(E_{0j}^\text{fl})^2=8\pi\hbar[n(\omega)+\frac{1}{2}]\text{Im}\{G_{jj}(\omega)\},
\end{equation}

\begin{equation}\label{sp}
(p_{0j}^\text{fl})^2=8\pi\hbar[n(\omega)+\frac{1}{2}]\text{Im}\{\alpha(\omega)\}.
\end{equation}
Inserting Eqs. (1,4-6) into the following integral \cite{Abajo-2010}:
\begin{equation*}
M=\frac{1}{4\pi^2}\int^{+\infty}_{-\infty}M_p(\omega)d\omega+\frac{1}{4\pi^2}\int^{+\infty}_{-\infty}M_E(\omega)d\omega,
\end{equation*}
we have a symmetrical expression:
\begin{align*}
M=&\frac{\hbar}{\pi}\int^{+\infty}_{-\infty}d\omega[n_1(\omega)+\frac{1}{2}]\text{Im}\{\alpha(\omega)\}[\text{Im}\{\bar{G}(\omega^-)-\text{Im}\{\bar{G}(\omega^+)]\\
+&\frac{\hbar}{\pi}\int^{+\infty}_{-\infty}d\omega[n_0(\omega)+\frac{1}{2}]\text{Im}\{\bar{G}(\omega)\}[\text{Im}\{\alpha(\omega^-)-\text{Im}\{\alpha(\omega^+)].
\end{align*}
Because each integral is convergent at the infinity frequencies for our cases, it is safe to change the integration limits and rewrite it as one compact expression:
\begin{equation}\label{storque}
M=-\frac{2\hbar}{\pi}\int^\infty_{-\infty}d\omega [n_1(\omega-\Omega)-n_0(\omega)]\text{Im}\{\alpha(\omega-\Omega)\}\text{Im}\{\bar{G}(\omega)\}.
\end{equation}

Following the same procedure (but in the rest frame), we can calculate the radiation power $P=-\textbf{E}\cdot \partial\textbf{p}/\partial{t}$ as
\begin{align}\label{sheat}
P_{1\to0}=&+\frac{2\hbar}{\pi}\int^\infty_{-\infty}d\omega \omega [n_1(\omega-\Omega)-n_0(\omega)]\text{Im}\{\alpha(\omega-\Omega)\}\text{Im}\{\bar{G}(\omega)\}\\
&+\frac{\hbar}{\pi}\int^\infty_{-\infty}d\omega \omega [n_1(\omega)-n_0(\omega)]\text{Im}\{\alpha(\omega)\}\text{Im}\{G_{zz}(\omega)\}.
\end{align}

\section{Supplemental Material: supplemental figures}

\begin{figure}[htb!]
\centering{\includegraphics[angle=0, width=8cm]{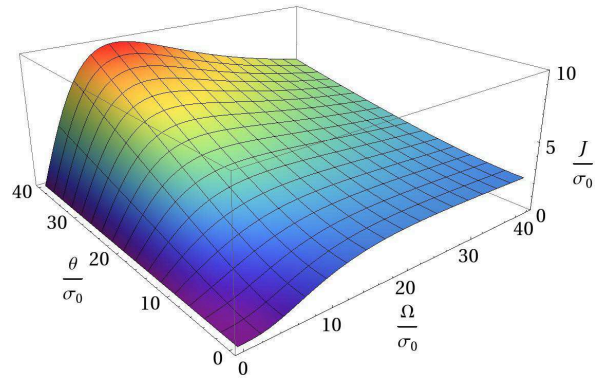}}
\caption{(color online) Three dimensional plot of $J$ versus the rotation frequency and temperature when the conductivity is fixed.}
\label{Fig:sstoppingtime}
\end{figure}

\begin{figure}[htb!]
\centering{\includegraphics[angle=0, width=8cm]{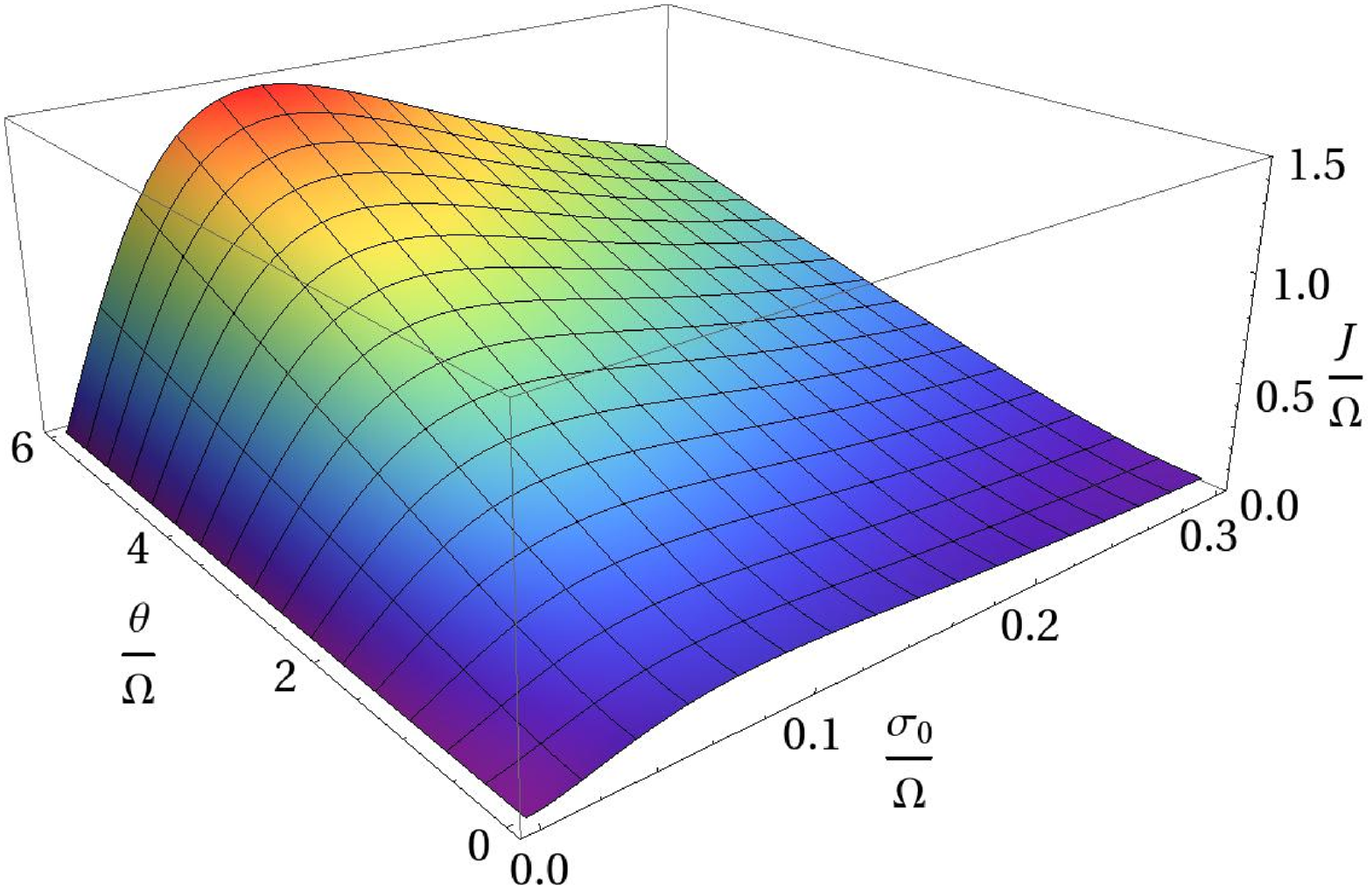}}
\caption{(color online) Three dimensional plot of $J$ versus the temperature and conductivity when the  rotation frequency is fixed.}
\label{Fig:sstoppingtime}
\end{figure}

\end{widetext}

\begin{thebibliography}{99}
\bibitem{Casimir-1948}
H. B. G. Casimir,
Proc. Kon. Nederl. Akad. Wet. \textbf{51}, 793 (1948).

\bibitem{Serry}
F.\,M. Serry, D. Walliser, and J. Maclay, J. Appl. Phys. \textbf{84}, 2501 (1998).
\bibitem{Buks}
E. Buks and M. L. Roukes,
Phys. Rev. B \textbf{63}, 033402 (2001).

\bibitem{Mostepanenko-2009}
G. L. Klimchitskaya, U. Mohideen, and V. M. Mostepanenko, Rev. Mod. Phys. \textbf{81}, 1827 (2009), and references therein.
\bibitem{Munday-2009}
J. N. Munday, Federico Capasso and V. A. Parsegian,
Nature \textbf{457}, 170 (2009).

\bibitem{Chan-2010}
Y. Bao, R. Gu\'erout, J. Lussange, A. Lambrecht, R. A. Cirelli, F. Klemens, W. M. Mansfield, C. S. Pai, and H. B. Chan,
Phys. Rev. Lett. \textbf{105}, 250402 (2010).
\bibitem{Lamoreaux-2011a}
A. O. Sushkov, W. J. Kim, D. A. R. Dalvit, and S. K. Lamoreaux, Nature Phys. \textbf{7}, 230 (2011).
\bibitem{Lamoreaux-2011b}
A. O. Sushkov, W. J. Kim, D. A. R. Dalvit, and S. K. Lamoreaux, Phys. Rev. Lett. \textbf{107}, 171101 (2011).



\bibitem{Johnson-2011}
A. W. Rodriguez, F. Capasso, and S. G. Johnson, Nat. Photon. {\bf 5}, 211 (2011), and references therein.
\bibitem{Zhao-2009}
R. Zhao, J. Zhou, Th. Koschny, E. N. Economou, and C. M. Soukoulis, Phys. Rev. Lett. \textbf{103}, 103602 (2009).






\bibitem{Pendry-1997}
J. B. Pendry, J. Phys. Condens. Matter {\bf 9}, 10301 (1997).
\bibitem{Pendry-1998}
J. B. Pendry, J. Mod. Opt. \textbf{45}, 2389 (1998).

\bibitem{Persson-2007}
A. I. Volokitin and B. N. J. Persson, Phys. Rev. Lett. \textbf{106}, 094502 (2011); Phys. Rev. B {\bf 78}, 155437 (2008); Rev. Mod. Phys. {\bf 79}, 1291 (2007), and reference therein.

\bibitem{Leonhardt-2009}
T. G. Philbin and U. Leonhardt, New J. Phys. {\bf 11}, 033035 (2009).

\bibitem{Pendry-2009}
J. B. Pendry, New J. Phys. {\bf 12}, 033028 (2010); U. Leonhardt, New J. Phys. {\bf 12}, 068001 (2010); J. B. Pendry, New J. Phys. {\bf 12}, 068002 (2010).
\bibitem{Persson-2011c}
A. I. Volokitin and B. N. J. Persson, New J. Phys. {\bf 13}, 068001 (2011); T. G. Philbin and U. Leonhardt, New J. Phys. {\bf 13}, 068002 (2011).

\bibitem{Golestanian-2007}
A. Ashourvan, M. F. Miri, and R. Golestanian,
Phys. Rev. Lett. {\bf 98}, 140801 (2007).

\bibitem{Abajo-2010}
A. Manjavacas and F. J. Garc\'ia de Abajo,  Phys. Rev. Lett. {\bf 105}, 113601 (2010); Phys. Rev. A {\bf 82}, 063827 (2010).

\bibitem{Greffet-2003}
K. Joulain, R. Carminati, J.-P. Mulet, and J.-J. Greffet,
Phys. Rev. B {\bf 68}, 245405 (2003).




\bibitem{SupplMat}
See Supplemental Material for the simple derivation for the torque and numerical results for $J$.

\bibitem{multiscattering}
Our expression for $\Gamma$ is like in Ref. [18], with the density of states $\rho_0$ substituted by $(3/2\pi^2\omega) \text{Im}\{\bar{G}\}$.


\bibitem{Jentschura-2012}
G. \L{}ach, M. DeKieviet, and U. D. Jentschura, Phys. Rev. Lett. {\bf 108}, 043005 (2012).

\bibitem{Kardar-2012}
M. F. Maghrebi, R. L. Jaffe, and M. Kardar, Phys. Rev. Lett. \textbf{108}, 230403 (2012).



\bibitem{Pendry-1999}
J. B. Pendry, J. Phys.: Condens. Matter {\bf 11}, 6621 (1999).
\bibitem{Irvin-1968}
S. M. Sze and J. C. Irvin, Solid-State Electron. \textbf{11}, 599 (1968).

\bibitem{Palik}
E. W. Palik, \textit{Handbook of Optical Constants of Solids I} (Academic, San Diego, CA, 1985).
\end{thebibliography}
\end{document}